\documentclass[aps, prd, reprint, longbibliography]{revtex4-1}
\usepackage{graphicx}
\usepackage{amssymb}
\usepackage{amsmath}
\usepackage{color}

\usepackage[normalem]{ulem}

\begin{document}

\title{Resolving anomalies in the critical exponents of FePt using finite-size scaling in magnetic fields}

\author{J Waters$^1$}
\email{J.M.Waters@soton.ac.uk}
\author{D Kramer$^1$}
\author{T J Sluckin$^2$}
\author{O Hovorka$^1$}
\affiliation{$^1$Engineering and Physical Sciences, University of Southampton, Southampton, SO17 1BJ, UK\\ $^2$School of Mathematical Sciences, University of Southampton, Southampton, SO17 1BJ, UK}

\date{\today}

\begin{abstract}
FePt is the primary material being considered for the development of information storage technologies based on heat-assisted magnetic recording (HAMR). A practical realization of HAMR requires understanding the high-temperature phase transition behavior of FePt, including critical exponents and Curie temperature distributions as the fundamental HAMR media design characteristics. The studies so far found a significant degree of variability in the values of critical exponents of FePt and remain controversial. Here we show that at the heart of this variability is the phase transition crossover phenomenon induced by two-ion anisotropy of FePt. Through Monte-Carlo simulations based on a realistic FePt effective Hamiltonian we demonstrate that in order to identify the critical exponents accurately, it is necessary to base the analysis on field-dependent magnetization data. We have developed a two-variable finite size scaling method that accounts for the field effect. Through the use of this method, we show unambiguously that true critical exponents of FePt are fully consistent with the three-dimensional Heisenberg universality class. \\

\end{abstract}

\maketitle

\section{Introduction}

Heat-assisted magnetic recording (HAMR) is a rapidly developing technology, designed to address the magnetic recording ``trilemma" \cite{weller2014hamr}.
Materials with high uniaxial anisotropy are required to increase thermal stability of magnetic grains, the most notable of these being FePt in the L$_1$0 phase. The high anisotropy of FePt is overcome during the recording process by laser heating the grains close to their Curie temperature $T_c$, which is dependent on the size of the FePt grains and their geometry. Understanding the dependence of $T_c$ on the finite size effects of magnetic grains in granular films is required for advancing HAMR technology.

The divergent behavior of thermodynamic quantities in materials near $T_c$ is described by critical exponents and universal scaling functions, which depend on the type and dimensionality of the material. Such critical exponents and universal scaling functions are used in a variety of studies of high-temperature magnetism. Examples include the models of thermalized dynamics based on the Landau-Lifshitz-Bloch equation \cite{atxitia2016fundamentals}, thermodynamic characterisation of materials near $T_c$ \cite[p.~80]{aharoni2000introduction}, or the experimental identification of $T_c$ distributions in magnetic granular media \cite{Waters2017, Oezelt2017}. Knowledge of accurate critical exponents is thus essential for a reliable high temperature quantification and optimization of materials for HAMR.

The critical exponents of FePt have been studied by fitting the power law behaviour to the magnetisation versus temperature $M(T)$ data \cite{Richter2017} or to the dependence of $T_c$ on the grain size $R$ \cite{rong2006size, Zhou2007}. Another widely used approach has to use finite size scaling analysis, rescaling $M(T)$ for grains of different $R$ so that they collapse onto a single curve \cite{hovorka2012curie, lyberatos2012size}. However, considerable variation of the critical exponents was found depending on the method used to identify them as illustrated in Table \ref{tab:literature}. Even small variations in the critical exponents can lead to large errors in the estimation of, for instance, the $T_c$ distributions \cite{hovorka2012curie}. It therefore becomes important to understand the reasons for these discrepancies and establish the values which should be used for the critical exponents.
\begin{table}[!b]
    \begin{tabular}{l || c | c | c}
        \hline \hline
        Critical paramaters \\ of FePt from literature &$\beta$ &$\nu$ &$T_c^b$ [K]\\
        \hline \hline
        Hovorka et. al. \cite{hovorka2012curie}
        &$0.33 \pm 0.10$ &$0.85 \pm 0.10$ &$677 \pm 11$ \\
        Lyberatos et. al. \cite{lyberatos2012size}
        &- &$1.06 \pm 0.06$ &$658 \pm 4$ \\
        Rong et al. \cite{rong2006size}
        &- &$0.67 \pm 0.11$ &$775 \pm 20$\\
        Zhao et. al. \cite{Zhou2007}
        &0.327 &0.631 &642.5 \\
        This work (Heisenberg)
        &$0.366 \pm 0.001$ &$0.72 \pm 0.17$ &$654 \pm 2$\\
        \hline \hline
    \end{tabular}
    \caption{\label{tab:literature}Values of the magnetisation and correlation length critical exponents $\beta$, $\nu$ and the bulk Curie temperature $T_c^b$ collected from the literature. The values obtained in this work using magnetic field dependent finite size scaling method fully agree with the values of the 3D Heisenberg model (Table \ref{tab:results}).}
\end{table}

\begin{figure*}
    \begin{center}
        \includegraphics[width=13cm]{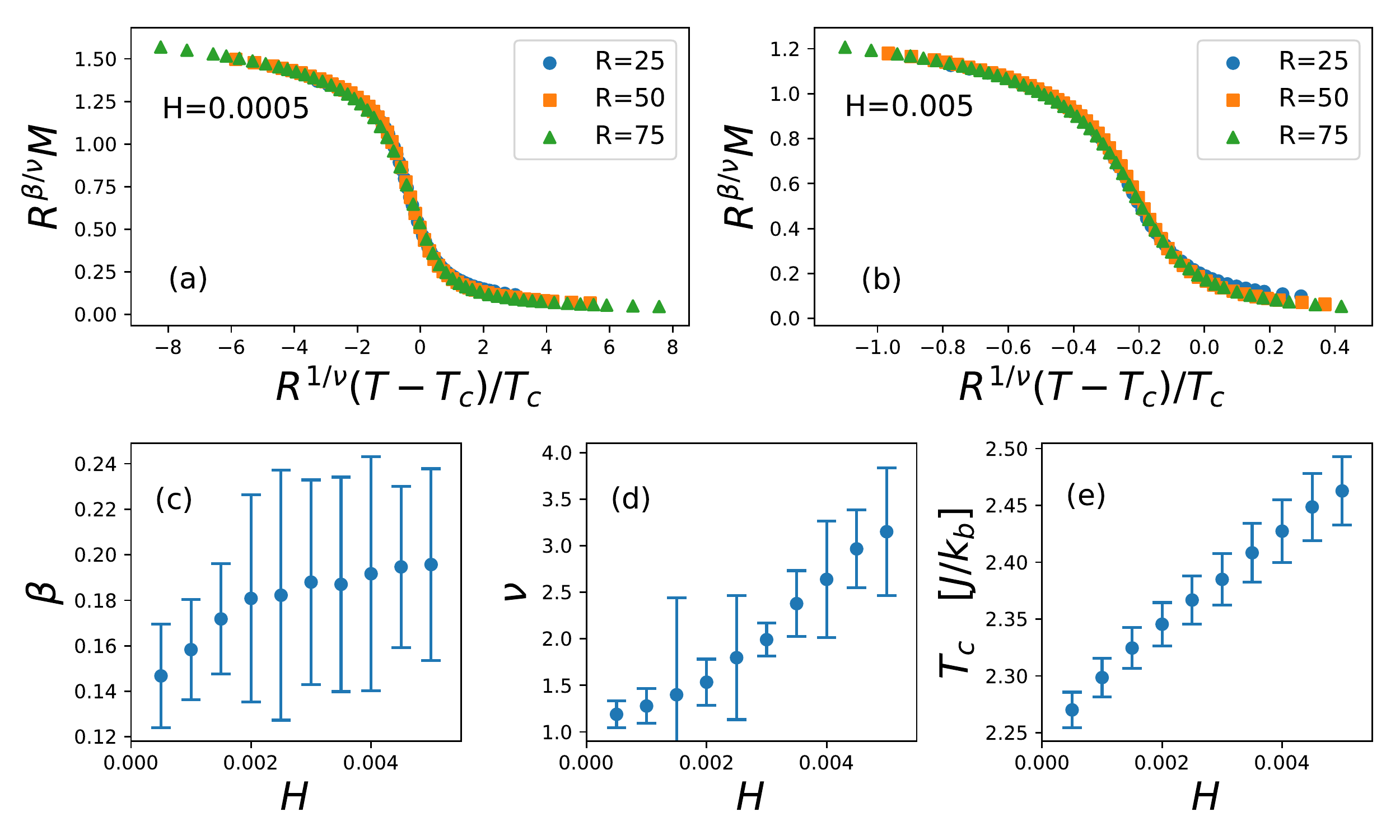}
        \caption{\label{fig:errorWithH} The results of finite-size scaling of Ising model data according to Eq. \ref{eq:1dfss}. (a) and (b) show the scaled curves when H = 0.0005 and H = 0.005 respectively. The resulting critical parameters (c) $\beta$, (d) $\nu$ and (e) $T_c^b$ which result from this method is plotted for a series of external field strengths. The errorbars represent the width of the minimum exponents at the 1\% level \cite{Bhattacharjee2001}.}
    \end{center}
\end{figure*}

In this work, we investigate this issue by employing large-scale Monte-Carlo simulations based on the FePt effective Hamiltonian obtained earlier from \emph{ab initio} calculations \cite{mryasov2005temperature}. We identify that the differences in the values of the critical exponents reported previously result from the phase transition crossover effect due to the presence of the two-ion anisotropy in FePt. Although being small in comparison to exchange interactions, the strength of the two-ion anisotropy is sufficient to conceal the information about the true critical exponents in the temperature region near $T_c$. We show that by generalising the finite size scaling analysis to consistently include the field-dependent magnetisation data allows to circumvent the crossover effect. In this way, we demonstrate unambiguously that, in the case where there is an external magnetic field, the critical exponents of FePt are fully consistent with the 3D Heisenberg model.

The article is organised as follows. In Section II, we develop the finite size scaling method, based upon the scaling of temperature and field-dependent magnetization data, $M(T, H)$, rather than $M(T)$. In Section III, we apply this technique to study the critical parameters of FePt and establish their relationship with the Heisenberg model. In Section IV, we discuss the crossover effects relevant to FePt and underline the dominant role of the two-ion anisotropy. In Section V, we conclude with recommendations for the critical exponents to use in future work.

%%%%%%%%%%%%%%%%%%%%%%%%%%%%%%%%%%%
\section{Two-Variable Finite Size Scaling}
%\subsection{Finite-Size Scaling of Field-Dependent Data}
%%%%%%%%%%%%%%%%%%%%%%%%%%%%%%%%%%%

In the present study we consider the finite size scaling (FSS) analysis as an established and robust technique for extracting the critical exponents in finite size simulations and small systems in general \cite{yeomans1992statistical, theorybook}. Considering magnetic grains of different diameters $R$, the FSS in its simplest form allows relating the temperature dependent $M(T)$ data for different $R$ to the magnetization and correlation critical exponents, $\beta$ and $\nu$, and the universal scaling function $\tilde M$ as:
\begin{equation}
    \label{eq:1dfss}
    M(T; R) = R^{-\beta/\nu}\tilde{M}\left(R^{1/\nu} t\right)
\end{equation}
where $t = (T - T_c^b)/ T_c^b$ is the reduced temperature, $T_c^b$ the Curie temperature of a bulk system, and $M$ represents the projection of the magnetization vector onto the $z$-axis. The critical exponents $\beta$ and $\nu$ can be identified by plotting the $M(T; R)$ data for different $R$ as  $R^{\beta/\nu}M$ versus $R^{1/\nu}t$, and systematically choosing the values of $\beta$, $\nu$ and $T_c^b$ until all data collapse onto a single universal curve representing $\tilde{M}$. A method convenient for producing such data collapses has been developed earlier \cite{Bhattacharjee2001}, and for completeness is reviewed here in Appendix A.

The FSS analysis based on Eq. (\ref{eq:1dfss}) is applicable only to $M(T; R)$ data in zero magnetic field, $H=0$. Non-zero magnetic fields introduce a rounding effect when $M$ no-longer approaches zero at the critical point but instead follows a power law behaviour $M\sim H^{1/\delta}$, where $\delta$ is the magnetic field critical exponent. To incorporate the magnetic field dependence of the data in the analysis it is necessary to generalise Eq. (\ref{eq:1dfss}) to include this field contribution, which results in the following two-variable FSS form:
\begin{equation}
    \label{eq:2dfss}
    M(T, H; R) = R^{-\beta/\nu}\tilde{M}\left(R^{1/\nu} t, R^{\beta\delta/\nu}H\right)
\end{equation}
The exponent $\delta$ is related to $\beta$ and $\nu$ through the so-called hyperscaling relation \cite{yeomans1992statistical}:
\begin{equation}
    \label{eq:hypsc}
    \delta = d\nu/\beta - 1
\end{equation}
with $d$ being the dimension of spin lattice, e.g. $d=3$ for three dimensional systems such as a grain in a magnetic hard disk. The form of Eq. \ref{eq:2dfss} suggests that the two-variable scaling approach is based on scaling the data seen as surfaces $R^{\beta/\nu}M(T, H; R)$ vs. $R^{1/\nu} t$ and $R^{\beta\delta/\nu}H$, and then tuning the values of $\beta, \nu, \delta$ and $T_c^b$ until a unique collapse onto the surface of $\tilde M$ is achieved. Details of our practical implementation of the two-variable scaling method are summarised in the Appendix A.
\begin{figure*}
    \includegraphics[width=15cm]{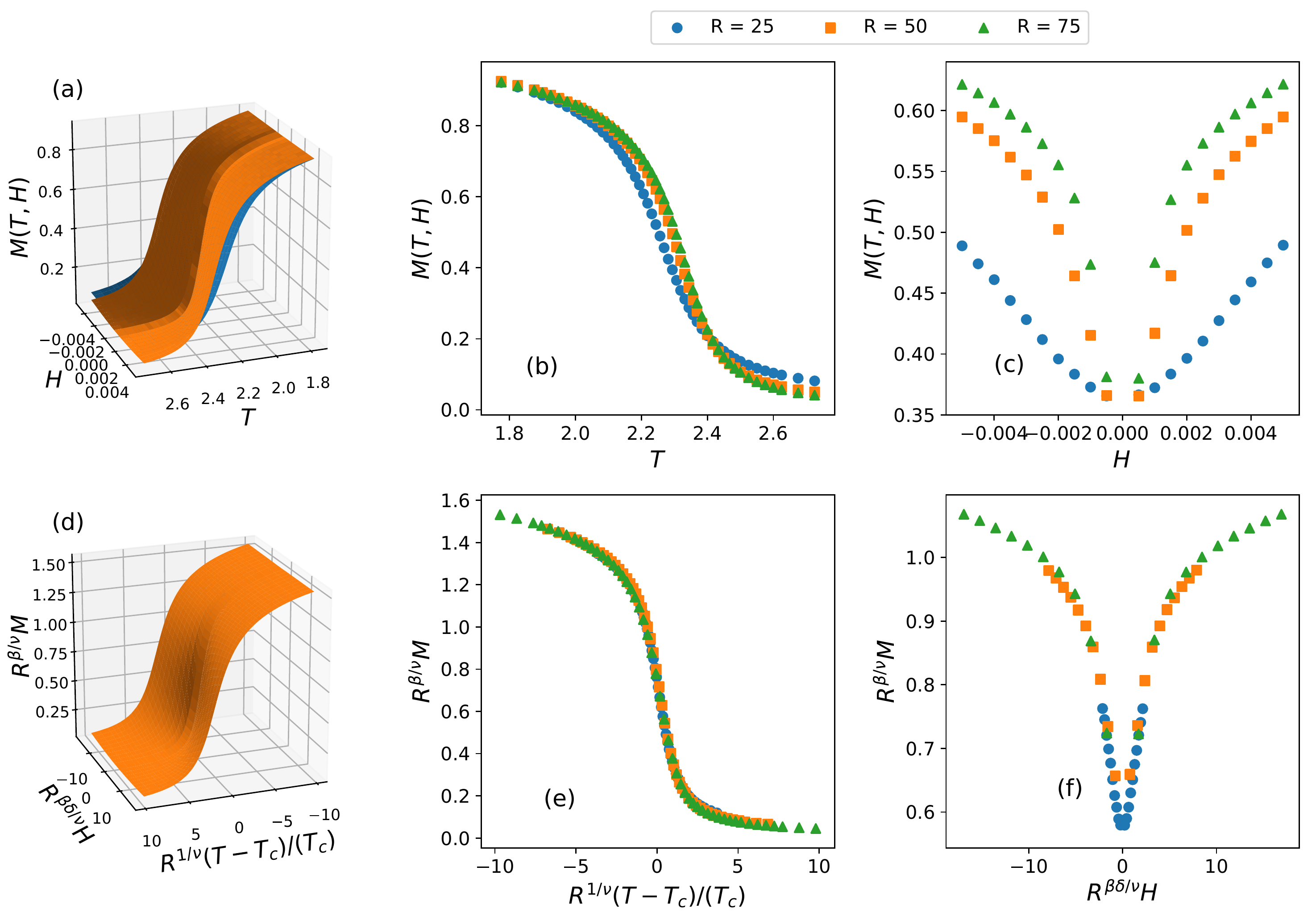}
    \caption{\label{fig:2dscale} (a) The magnetization surfaces in the $H$-$T$ plane corresponding to the Ising model. Slices through these surfaces are shown at (b) $H = 0.005$ and (c) $T = 2.25625$. (d) The same magnetization surfaces as shown in (a) after rescaling according to Eq. \ref{eq:2dfss}. Slices through these surfaces are shown at (e) $R^{\beta \delta / \nu}H = 1.871$ and (f) $R^{1/\nu} (T - T_c^b)/T_c^b = -0.029$.}
\end{figure*}

To validate the two-variable scaling approach based on Eq. (\ref{eq:2dfss}) we consider Metropolis Monte-Carlo simulations of a 2D Ising model, which is well established and rigorous analytical results for the critical exponents and $T_c$ are available for computational benchmarking.
The Hamiltonian of the field-dependent two-dimensional Ising model is $\mathcal{H} = - H\sum_i s_i - \sum_{ij}J_{ij}s_is_j$, where $H$ is the strength of the external magnetic field being applied to the system, $s_i$ is the $i$-th spin taking values $\pm 1$, and $J_{ij}>0$ is the ferromagnetic exchange interaction between $s_i$ and $s_j$ such that $J_{ij} = J = 1$ for neighboring spins and $J=0$ otherwise.
In the Monte-Carlo simulations, the magnetization was sampled for 3 different circular cuts of square lattices with radii $R=$ 25, 50 and 75 lattice spacings. Note that physical dimensions in this Ising model based test case are irrelevant. For each lattice size, the average magnetization was computed in the temperature range from $T = 1.775J/k_b$ to $2.725J/k_b$ and in the range of external field strengths of $H = -0.005J$ to $0.005J$. This encompasses the critical point at $H=0$ and $T_c=2.269J/k_b$. For each field strength, lattices were annealed from high to low temperature, with 10000 Monte Carlo sweeps (MCS) to equilibrate, then 51200 samples taken at intervals of 50 MCS.

\emph{Validation of one-variable FSS}. As a test case, the one-variable FSS analysis of only the $H=0$ data sets gave $\beta=0.14\pm 0.01$, $\nu = 1.05\pm0.08$ and $T_c^b = 2.265\pm 0.004$, which agree well with the theoretical values for the 2D Ising model, $\beta = 0.125$, $\nu = 1$ and $T_c^b = 2.269$ \cite{yeomans1992statistical}.
Next we apply the one-variable FSS to the 2D Ising model magnetization data obtained in non-zero magnetic field $H\ne0$. The obtained data collapses were successful for all considered field strengths, as suggested in Figs. \ref{fig:errorWithH}(a) and (b) for fields differing by a factor of 10. However, despite the successful data collapses, the obtained values of the critical parameters $\beta$, $\nu$ and $T_c^b$ tend to increase with $H$ as shown in Fig. \ref{fig:errorWithH}(c)-(d).
Thus the identified values of the critical exponents and $T_c^b$ appear to be incorrectly field-dependent despite the fact that the magnetic field is relatively weak and never greater than 0.5\% of the strength of the exchange field, and thus certainly not the dominant interaction. Analysis of the data using the zero-field ansatz in Eq. (\ref{eq:1dfss}) thus introduces a field-dependent bias of the critical exponents, which no longer represent the true critical exponents of the 2D Ising model. We have also attempted to extrapolate the field-dependent critical exponent data in Figs. \ref{fig:errorWithH}(c)-(e) to $H=0$ and obtained the critical parameters $\nu = 0.126$, $\beta = 0.988$ and $T_c^b= 2.256$, which agree with the theoretical values for the 2D Ising model surprisingly well. Thus the one-variable FSS combined with extrapolation to $h=0$ appears to be a valid technique for extracting the accurate values of the critical exponents. Unfortunately, as demonstrated below, due to the cross-over effects this procedure is not applicable to systems with more complex Hamiltonians containing contributions from magnetic anisotropies. This disqualifies this extrapolation approach for use with FePt.

\emph{Validation of two-variable FSS.} Fig. \ref{fig:2dscale}(a) shows the $M(T, H)$ surfaces obtained from the Monte-Carlo simulations of the 2D Ising model used above. Figs. \ref{fig:2dscale}(b)-(c) show the cuts through these data surfaces at specific values of the field and temperature. Fig. \ref{fig:2dscale}(d) shows the data collapse of these $M(T, H)$ data sets for all $R$ obtained by using the scaling procedure described in Appendix A. Figs. \ref{fig:2dscale}(e)-(f) demonstrate representative data cuts through the collapsed surface and suggest excellent scaling in both $H$ and $T$. The scaling procedure yielded $\beta = 0.136 \pm 0.018$, $\nu = 1.09 \pm 0.07$ and $T_c^b = 2.25 \pm 0.02$, which are in a very good agreement with the values of the 2D Ising model. Using Eq. \ref{eq:hypsc}, we find $\delta = 15.0 \pm 2.4$, in consistency with the analytical $\delta=15$. This demonstrates the validity of the two-variable FSS in the presence of magnetic fields, which will turn out to be essential for the analysis of the data of FePt below.

%%%%%%%%%%%%%%%%%%%%%%%%%%%%%%%%%%%
\section{Critical exponents of FePt}
\label{sec:method}
%\subsection{Monte Carlo Simulations}
%%%%%%%%%%%%%%%%%%%%%%%%%%%%%%%%%%%

To study the phase transition behavior of FePt we consider the following realistic classical effective spin Hamiltonian:
\begin{equation}
    \label{eq:FePtHamilField}
    \begin{split}
    \mathcal{H} = &-\sum_{ij}\left(J_{ij}\mathbf{s}_i \cdot \mathbf{s}_j + d_{ij}^{(2)}s_i^{z}s_j^{z}\right) \\
    &- \sum_i \left(d_i^{(0)}\left(s_i^{z}\right)^2 + \mu_{Fe}H s_i^{z}\right)
    \end{split}
\end{equation}
Here $\mu_{Fe} = 3.23\mu_B$ is the effective magnetic moment of an Fe atom with $\mu_B$ being the Bohr magneton, and $H$ is the strength of the magnetic field oriented along the $z$-axis. The Fe spins $\mathbf{s}_i$ are represented as Heisenberg spins with magnitude $|\mathbf{s}_i| = 1$. The effective exchange interaction $J_{ij}$ is not restricted to the nearest neighbor Fe spins, and while the contribution from the short-range interactions is stronger, the long-range contributions remain significant and cannot be truncated in the phase transition studies of FePt. Overall, this amounts to up to 1358 neighbor pairs for each spin in the crystal lattice. In addition, Eq. (\ref{eq:FePtHamilField}) contains a single-ion energy term with strength $d_i^{(0)}$ and a two-ion term with strength $d_{ij}^{(2)}$. These anisotropy terms introduce the tendency towards the spin alignment along the $z$-axis to maximise the $s_i^{z}$ spin component. The magnitude of these anisotropy terms is small and they act as a perturbation to the exchange energy term. For example, as discussed in more detail below, the strength of the two-ion anisotropy interaction is less than $1\%$ the strength of the exchange interaction.
The effective Hamiltonian given by Eq. (\ref{eq:FePtHamilField}) was determined by mapping the energy of FePt in the L$_1$0 phase to a tetragonal lattice of Fe atoms based on \emph{ab initio} calculations \cite{mryasov2005temperature}, and is now widely used for studying FePt for HAMR. 

\begin{figure}[!t]
    \includegraphics[width=8.5cm]{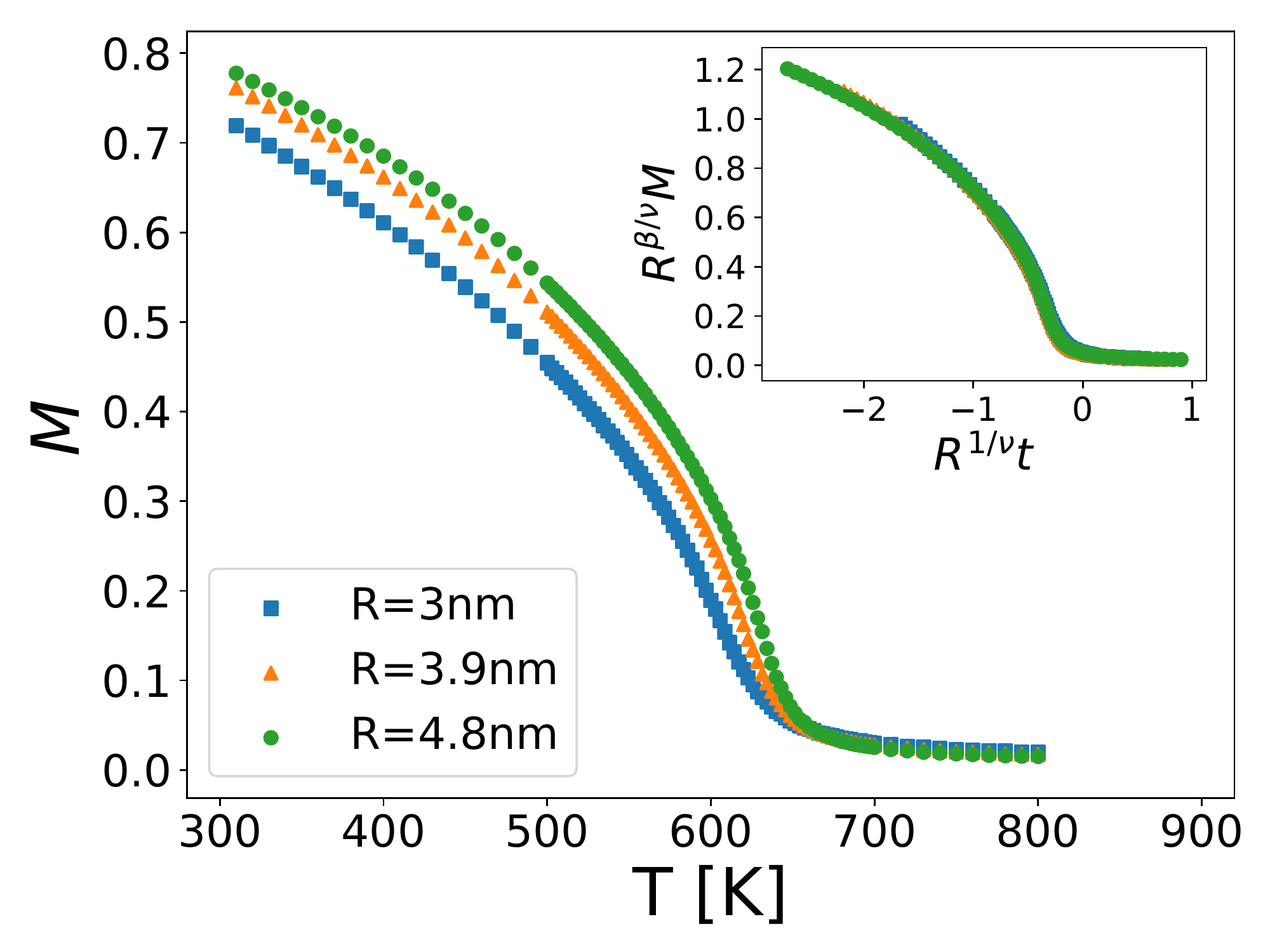}
    \caption{\label{fig:1dscale_FePt} The z-axis magnetization profiles of zero-field FePt data. Inset: The same magnetization profiles after rescaling according to Eq. \ref{eq:1dfss}.}
\end{figure}

To simulate the magnetic field and temperature dependent magnetisation data of the model Eq. (\ref{eq:FePtHamilField}) we use the Monte Carlo method introduced earlier. In simulations, we consider grains as spherical particles with radii $R$ in the range from 2 nm to 5 nm.
The magnetic field strengths normalised in the energy units as $\mu_{Fe}H$ were chosen in the range from 0 meV to 3 meV (0-16 Tesla), and the temperature  $T$ range from 310 K to 800 K.
The temperature dependent magnetisation curve at every field value was generated by applying the annealing protocol by decreasing the temperature self-consistently in small steps starting from 800 K. Equilibration time of 10000 MCS was applied for every field and temperature value, and 6048 samples were taken for averaging at intervals of 50 MCS.

%%%%%%%%%%%%%%%%%%%%%%%%%%%%%%%%%%%
%\section{Finite-size scaling of FePt}
%\label{sec:results}
%%%%%%%%%%%%%%%%%%%%%%%%%%%%%%%%%%%

\begin{figure*}[!t]
    \includegraphics[width=15cm]{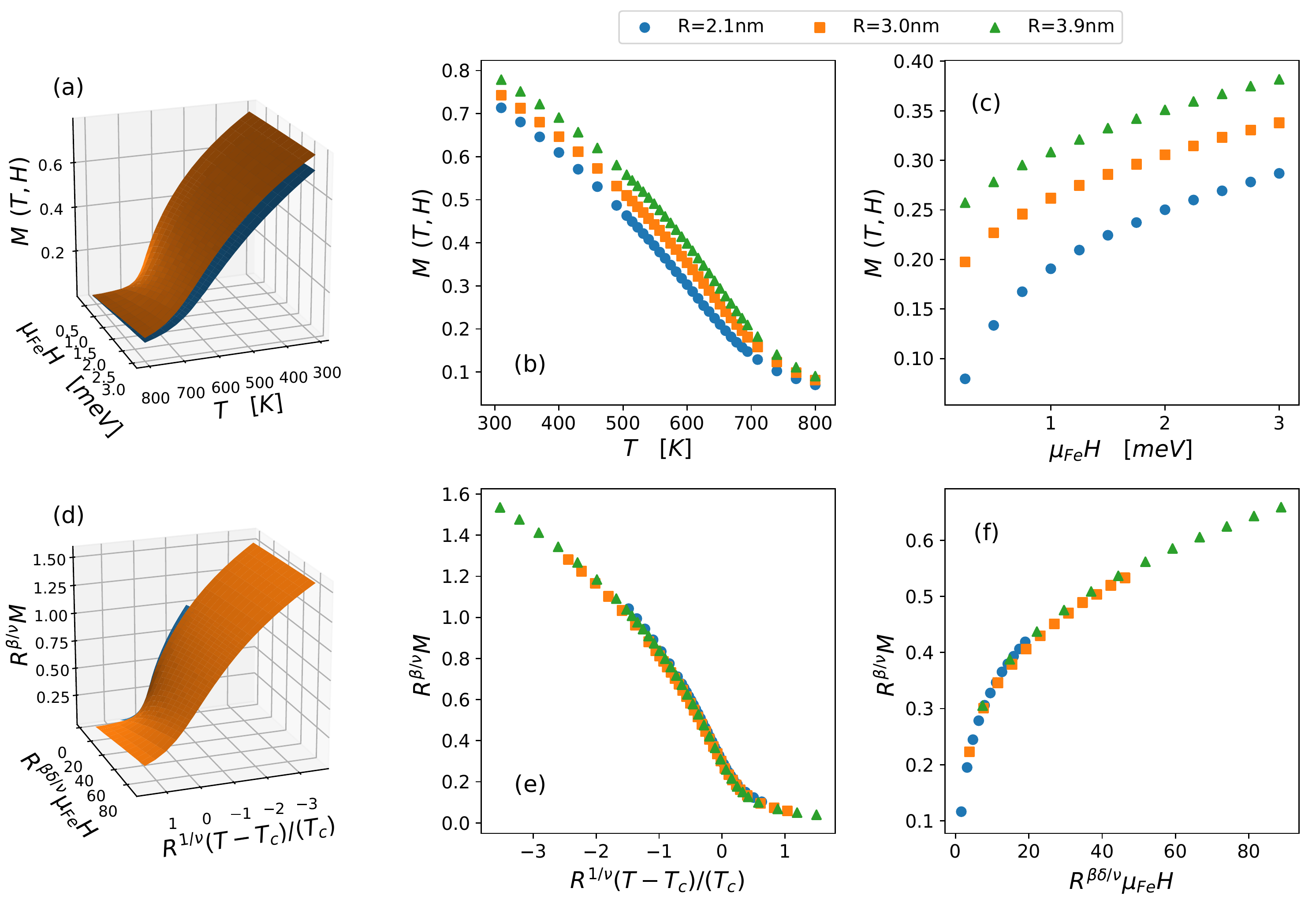}
    \caption{\label{fig:2dscale_FePt} (a) The magnetization surfaces in the $H$-$T$ plane corresponding to the field-dependent FePt data. Slices through these surfaces at (b) $\mu_{Fe}H = 3$meV and (c) $T = 608.57$K. (d) The same magnetization surfaces as shown in (a) after rescaling according to Eq. \ref{eq:2dfss}. Slices through these surfaces at (e) $R^{\beta \delta / \nu}\tilde{\mu}_{Fe}H = 18.88$ and (f) $R^{1/\nu} (T - T_c^b)/T_c^b = -0.20$.}
\end{figure*}

\emph{One-variable FSS.} We first identify the critical exponents of FePt using the one-variable FSS based on Eq. (\ref{eq:1dfss}). Fig. \ref{fig:1dscale_FePt} shows an example of the raw $M(T)$ data for particles of variable $R$ and corresponding to $H=0$. Here $M$ represents the $z$-component of the lattice averaged magnetisation vector.
The FSS analysis of these zero-field data led to excellent data collapse shown in the inset in Fig. \ref{fig:1dscale_FePt}. The values of the critical parameters were estimated as $\beta = 0.28 \pm 0.08$, $\nu = 0.94 \pm 0.15$ and $T_c^b = 678 \pm 15K$, and are reasonably consistent with our earlier work \cite{hovorka2012curie} as shown in Table \ref{tab:literature}. Any differences possibly attributable to the details of the data analysis, such as the density of data points or the choice of temperature range used in FSS. An over-reliance on temperatures close to $T_c$ can enhance the the effect of the crossover phenomena, as is discussed below.
We also attempted the one-variable FSS of the field- and temperature-dependent FePt data, similarly to the 2D Ising model test case in Fig. \ref{fig:errorWithH}. Unfortunately, we have not succeeded in producing meaningful data collapses and, therefore, the one-variable FSS based on Eq. (\ref{eq:1dfss}) did not allow us to identify the critical exponents from the FePt magnetisation data for non-zero magnetic fields.

\begin{table}[!b]
    \begin{tabular}{l || c | c | c}
        Critical Paramaters &$\beta$ &$\nu$ &$\delta$ \\ \hline \hline
        2D Ising model \\ \hline
        $H=0$ [Eq. (\ref{eq:1dfss})]&$0.14\pm0.01$ &$1.05\pm0.08$ &$-$ \\
        $H\ne0$ [Eq. (\ref{eq:2dfss})] &$0.136 \pm 0.018$ &$1.09 \pm 0.07$ &$15 \pm 0.02$ \\
        Analytic \cite{yeomans1992statistical} &0.125 &1 &$15$ \\
 \hline \hline
        FePt \\ \hline
        $H=0$ [Eq. (\ref{eq:1dfss})]&$0.28 \pm 0.08$ &$0.94 \pm 0.15$ &$-$ \\
        $H\ne0$ [Eq. (\ref{eq:2dfss})]&$0.366 \pm 0.001$ &$0.72 \pm 0.17$ &$4.9\pm0.9$ \\
        3D Heisenberg \cite{yeomans1992statistical}&0.36 &0.71 &$4.8$ \\ \hline \hline
    \end{tabular}
    \caption{\label{tab:results} The results of FSS analysis performed in this work and the expected theoretical values. For comparison, the summary of the data from previous studies is given in Table \ref{tab:literature}.}
\end{table}

\emph{Two-variable FSS.} Fig. \ref{fig:2dscale_FePt}(a) shows the full set of the FePt magnetisation data computed for different $T$, $H$, and $R$.  Figs. \ref{fig:2dscale_FePt}(b)-(c) show slices through the data surfaces at specific values of $H$ and $T$.
The corresponding data collapse obtained by using the two-variable FSS ansatz Eq. (\ref{eq:2dfss}) is shown in Fig. \ref{fig:2dscale_FePt}(d). The representative collapsed data cuts are shown in Fig. \ref{fig:2dscale_FePt}(e)-(f) and confirm that good scaling has been achieved.
The critical parameters obtained from the two-variable FSS analysis were $\beta = 0.366 \pm 0.001$, $\nu = 0.72 \pm 0.17$ and $T_c^b = 654 \pm 2K$, and the magnetic field exponent as found from Eq. \ref{eq:hypsc} was $\delta = 4.9 \pm 0.9$. These results are in a very good agreement with the known exponents for the 3D Heisenberg model for which $\beta = 0.36$, $\nu = 0.71$ and $\delta = 4.8$ \cite{yeomans1992statistical}.

Table \ref{tab:results} summarises the results of the FSS analysis of the zero and non-zero-field two-dimensional Ising model and FePt magnetization data studied in this work.
The critical exponents of the Ising model obtained by one-variable FSS of the $H=0$ data and two-variable FSS of the $H\ne0$ data are both consistent with the theoretical values.
For FePt, the two-variable FSS of the $M(T, H)$ data gave critical exponents consistent with the Heisenberg model universality class. However, the one-variable FSS of the zero-field FePt data was not consistent with the Heisenberg model. The question arises if the origin of this inconsistency, absent in the Ising model case, is inherent to the chosen data analysis or if it is of fundamental nature.
Below we argue that it is the result of the presence of anisotropy terms in the effective Hamiltonian Eq. (\ref{eq:FePtHamilField}), leading to phase transition crossover effects and complicating the identification of accurate values of the critical exponents from the zero-field data.

%%%%%%%%%%%%%%%%%%%%%%%%%%%%%%%%%%%
%\section{Discussion}
\section{Crossover effect}
\label{sec:disc}
%%%%%%%%%%%%%%%%%%%%%%%%%%%%%%%%%%%

\begin{figure}[!t]
    \includegraphics[width=8.5cm]{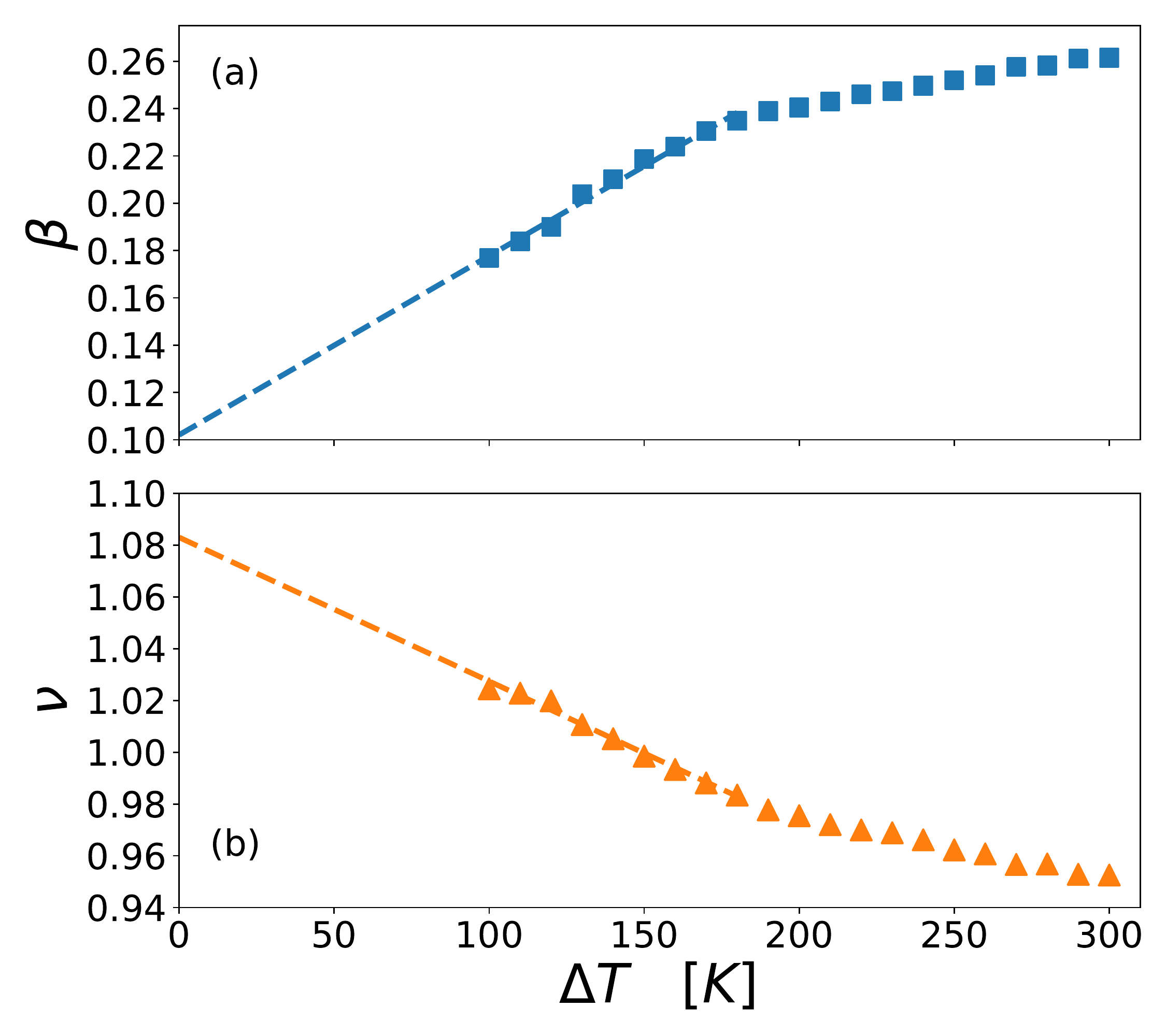}
    \caption{\label{fig:delT_FePt} The values of (a) $\beta$ and (b) $\nu$ found from zero-field scaling of FePt magnetization curves in zero-field (according to Eq. \ref{eq:1dfss}) using data in the range $[T_c^b - \Delta T, T_c^b + \Delta T]$, using increasing values of $\Delta T$. The dashed lines shows a naive extrapolation to $\Delta T = 0$.}
\end{figure}

The most common sources of the crossover behaviour during phase transitions are (i) small residual fields acting on the system, (ii) presence of  weak anisotropies and long-range interactions, (iii) effect of disorder, (iv) and finite size effects \cite{theorybook}. We focus on analysing the case (ii) only, given that simulations allow setting well-controlled external field conditions, unlike experiments, and thus ruling out the residual files in (i); atomic disorder is not considered and, therefore, its effects (iii) are irrelevant in our simulations; and the finite system size effects (iv) are accounted for by choosing the analysis method based on the FSS ansatz Eqs. (\ref{eq:1dfss}) or (\ref{eq:2dfss}), rather than naive power law fits.

\emph{Effect of anisotropy.} It is useful to first estimate the magnitude of contributions of the individual terms in the effective Hamiltonian Eq. (\ref{eq:FePtHamilField}) in the critical temperature region. The effective field acting on a spin $i$ expressed in the mean-field approximation reads $\mu_\mathrm{Fe}\mathbf{H}_{\mathrm{eff}, i} = -\partial\mathcal{H}_\mathrm{mf}/\partial\langle\mathbf{s}_i\rangle$, where $\langle\mathbf{s}_i\rangle$ is the expectation value of the spin $\mathbf{s}_i$ consistent with the equilibrium Boltzmann statistics. According to Eq. (\ref{eq:FePtHamilField}), the effective field $\mu_\mathrm{Fe}\mathbf{H}_{\mathrm{eff}, i}$ can be written as a sum of contributions:
\begin{equation}
\sum_j J_{ij}\langle\mathbf{s}_j\rangle + \sum_jd_{ij}^{(2)}\langle {s_j^z}\rangle  + d_i^{(0)}\langle {s_i^z}\rangle + \mu_\mathrm{Fe}H
\end{equation}
where the first term defines the local exchange field, the second and third terms are the local two-ion and single-ion anisotropy fields, and the last term is the external field. All fields are in the units of energy.
Taking $|\langle\mathbf{s}_i\rangle| = 0.05$ as a representative value of magnetisation in $H=0$ at 650 K near the phase transition point (Fig. \ref{fig:1dscale_FePt}) we estimate the exchange, two-ion and single-ion anisotropy to be approximately 11 meV, $0.15$ meV and $8.45\times 10^{-5}$ meV, respectively.
In the following analysis, the single-ion anisotropy contribution can therefore be neglected, and the two-ion anisotropy taken as a small perturbation to the exchange energy term.

\begin{figure}[!t]
    \includegraphics[width=8.5cm]{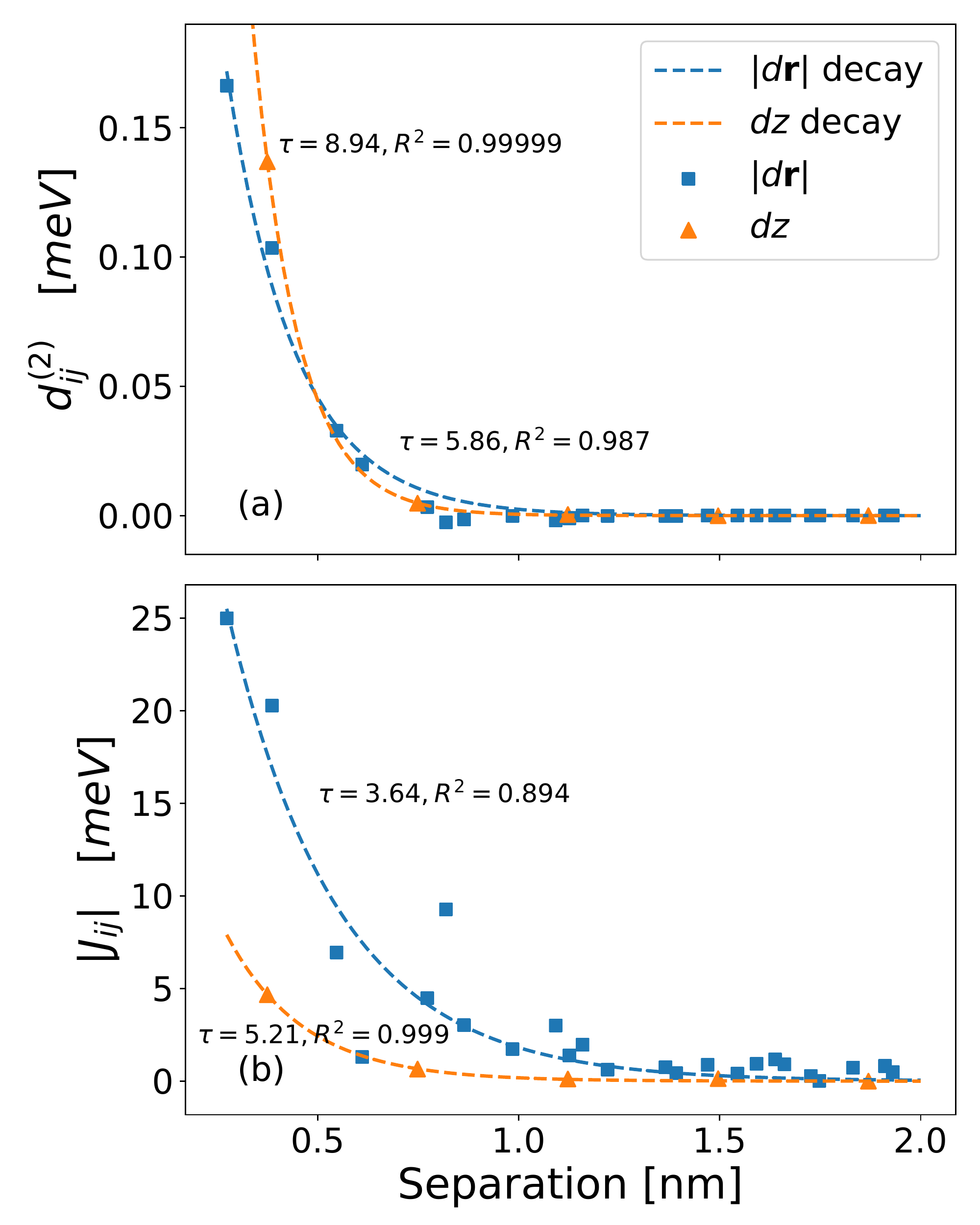}
    \caption{\label{fig:FePt_ani} The strength of the interactions of Fe spins based on their separation for in-plane particles (blue) and out-of-plane particles (orange). (a) shows the strength of the two-ion anisotropy and (b) shows the strength of the effective exchange.}
\end{figure}

The anisotropy-induced crossover effect can be understood by the following qualitative argument \cite{theorybook} applied to the zero-field Heisenberg Hamiltonian with the two-ion anisotropy acting as a small perturbation. The scaling form of the free energy of such a system reads $f_s = |t|^{2-\alpha}\tilde F(\langle d_{ij}^{(2)}\rangle |t|^{-\phi}$), where $\tilde F$ is the universal scaling function for the free energy, $t = (T-T_c^b)/T_c^b$ is the reduced temperature as before, $\phi > 0$, and $\langle d_{ij}^{(2)}\rangle$ is an estimate of the characteristic two-ion anisotropy strength such as an average over all $d_{ij}^{(2)}$, for example. It can be shown that asymptotically, the function $\tilde F(x)$ for $|x|<< 1$ represents the effective Heisenberg exchange-dominated system, while for $|x|>>1$ the behaviour is Ising-like \cite{theorybook}. Which of the two asymptotic regimes will be attained by the system at a given temperature depends on the value of $x =\langle d_{ij}^{(2)}\rangle|t|^{-\phi}$. For temperatures sufficiently far from $T_c$ one expects $|t|$ to be large and $\langle d_{ij}^{(2)}\rangle|t|^{-\phi}\rightarrow 0$, given that $\phi$ is positive, and thus the behavior to be Heisenberg-like. However, for measurements close to $T_c$, the reduced temperature $t\rightarrow 0$ and $\langle d_{ij}^{(2)}\rangle|t|^{-\phi}\rightarrow\infty$, leading to an Ising-like asymptotic regime. The results of the analysis will therefore strongly depend on the temperature interval selected for the analysis.

To see this, we performed one-variable FSS analysis similar to Fig. \ref{fig:1dscale_FePt} but with a restricted range of temperatures from $T_c^b - \Delta T$ to $T_c^b + \Delta T$, and varying the $\Delta T$.
Fig. \ref{fig:delT_FePt} shows the exponents $\beta$ and $\nu$ identified in this way. As $\Delta T$ decreases there is a change of the trend near $T=180$ K. Exponent values below about 90 K could not be obtained due to the insufficient data points available in the now narrowly restricted temperature interval, and instead we used naive linear extrapolation to estimate the exponents for $\Delta T=0$.
The extrapolation procedure gave $\beta \approx 0.1$ and $\nu \approx 1.08$, which are, according to Table \ref{tab:results}, indeed close to the critical exponents for the 2D Ising model.
Thus, even small anisotropic effects can completely alter the estimated value of the critical exponents. Note that the values of the critical exponents in the high $\Delta T$ limit in Fig. \ref{fig:delT_FePt} are not expected to have a physical meaning due to the mixed contributions from both asymptotic regimes.

The anisotropy-induced crossover behavior can be overcome by performing the data analysis in the non-zero applied magnetic field, stronger than the anisotropy field but sufficiently small to avoid driving the system away from the phase transition region. For our two-variable FSS analysis we have applied magnetic fields in the interval 0 - 3 meV (0 - 16 Tesla), which dominate the two-ion anisotropy field (0.15 meV) and, given that the data scaling is excellent, still within the phase transition region.
Given that the two-variable scaling analysis systematically incorporates the effects of magnetic fields in the phase transition region, this allowed resolving the crossover effect and establishing correct critical exponents in the universality class of the 3D Heisenberg model.
Thus the two-variable FSS based on Eq. (\ref{eq:2dfss}) is essential in the identification of the correct exponents. Also, Fig. \ref{fig:2dscale_FePt} (f) shows that the data collapse is good through the entire range of magnetic fields. Therefore, in practice, smaller fields may be all that is necessary for the analysis, such as are achievable by a standard laboratory magnetometry.

\emph{Effect of long-range interactions.} The long-range nature of the interaction and two-ion anisotropy energy terms as a potential contributing factor to the crossover behaviour can be understood based on the following standard argument \cite{theorybook, Fisher1972}. If the interaction strength $J_{ij}$ between spins at locations $\mathbf r_i$ and $\mathbf r_j$ along a given crystallographic direction decays as a function of their separation $r_{ij} = |\mathbf r_i-\mathbf r_j|$ exponentially fast, i.e. if they can be fitted to a function $f_\mathrm{exp}(r_{ij}) = A\exp\left(-r_{ij} / {\tau}\right)$, then the interaction can be considered as short range and not contributing to the crossover. Given that the interaction and the two-ion anisotropy profiles in the effective Hamiltonian of FePt in Eq. (\ref{eq:FePtHamilField}) have directional variability, we apply this argument to in-plane and out-of-plane interaction and anisotropy pairwise coupling distributions. Fig. \ref{fig:FePt_ani}(a) shows the plots of in-plane and out-of-plane two-ion anisotropy couplings as a function of the spin-spin separation $r_{ij}$. The raw data are fitted very well by exponential function $f_\mathrm{exp}$, confirming the short range character of the two-ion anisotropy. Similarly, Fig. \ref{fig:FePt_ani}(b) shows the dependence of in-plane and out-of-plane of the interaction coupling strength $|J_{ij}|$ vs. $r_{ij}$, where the modulus is taken because of the oscillatory character of the $J_{ij}$ with increasing spin-spin separation. Once more, the functional dependence is represented very well by the exponential decay $f_\mathrm{exp}$, and thus the interactions in Eq. (\ref{eq:FePtHamilField}) can be considered as short range. Overall, this effectively rules out the long-range interactions as driving the crossover behavior and suggests that the two-ion anisotropy is the primary contributing factor.

%%%%%%%%%%%%%%%%%%%%%%%%%%%%%%%%%%%
\section{Conclusions}
%%%%%%%%%%%%%%%%%%%%%%%%%%%%%%%%%%%

In summary, the present study resolves the issue of the variability of critical exponents which have been reported in earlier research (Table \ref{tab:literature}) and shows that, in the presence of a magnetic field, the critical exponents of FePt are fully consistent with those of the three-dimensional Heisenberg model. The discrepancies between the various groups exponents in Table \ref{tab:literature} can be attributed to the phase transition crossover behavior, which renders the identification of critical exponents highly sensitive and dependent on the details of the analysis. In particular, our study suggests that it is not possible to identify accurate values of critical exponents of FePt based on zero-field magnetisation data, and instead field-dependent data need to be included in the analysis. The origin of the crossover behaviour is in the presence of the two-ion anisotropy term in the effective Hamiltonian of FePt (Eq. \ref{eq:FePtHamilField}), which acts as perturbation to exchange energy, and can become relevant in the phase transition region close to $T_c$. While this crossover effect dominates the zero-field magnetization data, it can be counter-balanced by applying the external magnetic field.

Our study has broad implications for the HAMR technology design. Since significant magnetic fields are used in the writing process \cite{Lyberatos2015}, it suggests that in order to quantify the $T_c$ distributions for quality control of the recording media, the Heisenberg critical exponents can be used in the design directly. This avoids the need for further independent identification of critical exponents anytime new experimental realizations of FePt-based HAMR media become available. At the fundamental level, our study suggests that the crossover behavior of FePt during phase transitions is governed by the two-ion anisotropy energy, rather than the long-range nature of exchange interactions, and can be circumvented through application of magnetic fields accessible by using standard laboratory magnetometry.

\section{Acknowledgements}

JW acknowledges financial support from the EPSRC Centre for Doctoral Training grant EP/L006766/1. This worked used the ARCHER UK National Supercomputing Service (http://www.archer.ac.uk), accessed through the UK HEC Materials Chemistry Consortium, which is funded by EPSRC (EP/L000202).This work also made use of the IRIDIS High Performance Computing Facility, and associated support services at the University of Southampton.

We also thank Matthew Ellis for providing the parameterization data for the FePt Hamiltonian.

\bibliography{Bibliography}

\section*{Appendix A}

The method of data collapse which is proposed in this manuscript is an adaptation of the automatic collapse proposed by Bhatacharjee and Seno \cite{Bhattacharjee2001}. A measure is proposed for the success of the collapse of various curves as a function of the critical exponents $\beta$ and $\nu$, which can be minimized in order to find the optimum values of the critical exponents. Following this method, if the function $\tilde{M}$ is known, then when rescaling a set of points $M(t)$ according to Eq. (\ref{eq:1dfss}), the total residual $r$ can be given as:
\begin{equation}
    r = \frac{1}{N} \sum \left| R^{\beta/\nu} M(t, R) - \tilde{M} \left(R^{1/\nu} t\right) \right|
\end{equation}
In general however, $\tilde{M}$ is unknown. Instead, interpolated values from another rescaled curve can be used in place of $\tilde{M}$. If it is assumed then, that a perfect collapse of several curves is one where the total residual is zero, then the best collapse of an imperfect dataset can be considered to be the one which minimizes $r$, i.e. minimizing the quantity $P_b$ \cite{Bhattacharjee2001}, where:
\begin{equation}
    \label{eq:oldscaling}
    P_b = \frac{1}{N_{ol}}\left[\sum_p \sum_{j\neq p} \sum_{i, ol} \left| R_j^{\beta/\nu} M_j(t_i) - R_p^{\beta/\nu} M_p^*(t_i^*) \right|^{q} \right]^{1/q}
\end{equation}
where $M_j$ is the set of samples of $M$ associated with system size $R_j$, $M_j^*$ is an interpolating function of the dataset $M_j$, $t_{i}$ is the $i$-th member of a dataset in the variable $t$, and $t_{i}^*$ is the value of $t_i$, rescaled from system size $R_p$ to $R_j$, i.e. $t_{i}^* = (R_j / R_p)^{1/\nu} t_{i}$. The value $N_{ol}$ is equal to the total number of datapoints which overlap between all pairs of curves with different $R$ after rescaling. Likewise, the sum over the points $t_{i}$ is only performed for those points where $t_{i}^*$ is within the extent of the interpolated curve $M_p^*$. Choosing only these points avoids errors due to extrapolation. The function $P_b$ is minimized through variation of $\nu$, $\beta$ and $T_c^b$, identifying them for that specific material. The value $q$, introduced in Eq. \ref{eq:oldscaling}, should have no effect on the results of the collapse, however it may effect the convergence of the method in some cases. For this work, $q$ was chosen to be equal to 1.

The errors on critical parameters which are shown in this work are the width of the minimum value of $P_b$ to a given level. This can be calculated as:
\begin{equation}
    \begin{split}
        \Delta \beta &= \eta \beta \left(2 \ln \frac{P_b(\beta \pm \eta \beta, \nu, T_c^b)}{P_b(\beta, \nu, T_c^b)}\right)^{-1/2} \\
        \Delta \nu &= \eta \nu \left(2 \ln \frac{P_b(\beta, \nu \pm \eta \nu, T_c^b)}{P_b(\beta, \nu, T_c^b)}\right)^{-1/2} \\
        \Delta T_c^b &= \eta T_c^b \left(2 \ln \frac{P_b(\beta, \nu, T_c^b \pm \eta T_c^b)}{P_b(\beta, \nu, T_c^b)}\right)^{-1/2}
    \end{split}
\end{equation}
where $\Delta \beta$, $\Delta \nu$ and $\Delta T_c^b$ are the widths of $\beta$, $\nu$ and $T_c^b$ to the level of $\eta$. i.e. At the 1\% level, $\eta=0.01$.

A modification to Eq. \ref{eq:oldscaling} is required in order to take into account the scaling with magnetic field strength, however, the minimization principle remains the same. The minimisation quantity $P_b$ is redefined so as to take into account the two-variable FSS form given in Eq. (\ref{eq:2dfss}):
\begin{equation}
    \label{eq:newscaling}
    P_b = \frac{1}{N_{ol}}\left[\sum_p \sum_{j\neq p} \sum_{i, ol} \left| R_j^{\beta/\nu} M_j(\mathbf{x_i}) - R_p^{\beta/\nu} M_p^*(\mathbf{x_i^*}) \right|^{q} \right]^{1/q}
\end{equation}
where $\mathbf{x_i}$ is the set of scaling variables $\{t_i, H_i\}$ for the $i$-th datapoint and $\mathbf{x_i^*}$ is the rescaled dataset $\{(R_j / R_p)^{1/\nu} t_{i}, (R_j / R_p)^{\beta\delta/\nu} H_{i}\}$. In Eq. \ref{eq:newscaling}, the function $M_j$ and the interpolating function $M^*_p$ are now surfaces rather than curves. The sum over $\mathbf{x_i}$ is performed only for those points where $\mathbf{x_i^*}$ lies within the extent of the interpolated surface $M_p^*$.

\end{document}